# The Titicaca basin: a paradigmatic region for multidisciplinary studies

Amelia Carolina Sparavigna
Dipartimento di Fisica
Politecnico di Torino, Italy

**Abstract**
We propose a survey of the literature about archaeological and geophysical researches and on the paleoclimatology of the basin of Titicaca Lake, in connection with the agricultural system of the raised fields. Here again we discuss the use of Google Maps and satellite imagery as fundamental tools to understand this ancient agricultural system and its role in the history of pre-Incaic cultures.

Sitting 3,811 m above sea level, Lake Titicaca is in a basin high in the Andes on the border of Peru and Bolivia. The western part of the lake lies within the Puno Region of Peru, and the eastern side is located in the Bolivian La Paz Department. Both regions have terraced hills and plains covered with raised fields, representing the remains of a huge agricultural system. Near the lake, in Bolivia, we find the well-known ruins of Tiwanaku.
Actually, the Titicaca basin can be a paradigmatic region for the growth of several multidisciplinary studies. There are many interesting researches in archaeology and anthropology, geophysical analysis and remote sensing investigations: all these studies need to be compared to answer questions that are still open on the history of this area. Here we discuss in particular the ancient agricultural system of the raised fields that can be easily and freely observed with Google Maps.
From the point of view of archaeological/anthropological studies, besides of course the researches on the Tiwanaku area with its monumental remains, the "raised fields" are quite important. This system of fields is an old technique of soil and water management, consisting of a series of earthworks on which crops can grow, surrounded by water canals. A known benefit of this system is the frost mitigation during the night, avoiding the damage of crops.
An interesting anthropological paper was recently published, approaching the "raised fields" of Titicaca lake in the framework of the organization of ancient intensive farming, comparing the "top-down" and "bottom-up" perspectives [1]. The "top-down" approach is that considering the development of intensive farming and its social organization as attributed to the rule action of a centralized government. The "bottom-up" instead is viewing an intensive farming as the incremental work of local communities or kin-based groups. The authors analysed in particular the Katari Valley [1], near the lake in Bolivia, on a long-term perspective covering 2500 years. They determined that the rural organization changed greatly over time in relation to changing socio-political conditions: that is the local communities played dynamic roles in the development and organization of raised field farming, but growth and ultimate recession were locked to the consolidation and decline of the Tiwanaku state. In fact, the authors arrived to the interesting conclusion that the top-down/bottom-up dichotomy is overdrawn.
In [1], we read that the top-down interpretation has roots in a Western social thought, commonly attributing the development of large-scale farming and irrigation systems to centralized governments and nascent states. And in fact, from this point of view, it is paradigmatic the Roman Empire, with its road and hydraulic engineers, where the central government organized the construction and maintenance of roads, aqueducts and also entertainment monumental places. Recent alternative perspectives emphasize that cultural creativity and political power was also the product of local groups, not only the product of central governments: that is, a large-scale economic production can be yielded by local kin-based groups, where elites or leaders coordinated such activities [1-3].

For what concerns the raised fields, "top-down" versus "bottom-up" interpretations have been proposed [4-6]. Proponents of both interpretations argue that intensive production was highly effective in the Titicaca region: in the top-down interpretations, intensive production was driven by the impetus of a centralized state government, whereas in bottom-up interpretations, it was locally developed and organized.

As reported in Ref.1, "determining who developed and managed intensive production in any specific case and with what technologies and resources requires rigorous interdisciplinary collaboration and empirical research". It is clear that only multidisciplinary researches will be able to solve the open questions about Titicaca, that, as reported in [1], are the following: When were raised fields built and by whom? When and why were they abandoned? Did raised fields require state management, or were they the exclusive domain of local communities?

In [1], the researchers focussed on an area in Bolivia on a long-term (ca. 2500 years) characterization of rural society and production dating from the emergence of complex societies until European colonization. The intensive raised field system adapted its predominant production to economic demands and socio-political conditions [1]. Based on research in the northwest Titicaca basin, near Huatta, Peru, Erickson [1,5] proposed that the raised field agriculture developed out of the knowledge and skills of communities and kin-based social groups, or "ayllus", who survived the subjugation by Andean states. Erickson ([1,7], p. 315) points out that, raised field agriculture "differs... in that there is no necessarily inherent need for large-scale cooperation, in the construction, use, nor maintenance of the system" and concluded that "to suppose that raised field farming could only be planned, executed and maintained by the highly centralized state is to disregard the rich agricultural knowledge and organizational potential of the Andean farmer." ([1,5], p. 413) Of course, other researchers disagree with this conclusion.

As previously told, among the open questions, it remains that on when the raised field system was firstly developed. The debate then centers on the relationship between settlements and raised fields and on the chronology of raised field construction and use. Erickson ([1,7], pp. 377–380) proposes two phases of raised-field construction in the north-western basin of Titicaca: First Phase, dating to the Early and Middle Formative periods (1500–200 BC); and Second Phase, dating to the post-Tiwanaku period [1,7]. In this chronological scheme, raised fields fell into disuse during the intervening Tiwanaku culture. It seems then that the period of growth of the raised fields in Peru is different from that of the opposite region near Tiwanaku, as in a counter-phase phenomenon.

From the analysis of the Google Maps, we have clearly observed that the system of raised fields, canals and artificial ponds in the Peruvian region of Titicaca contains peculiar area where the earthworks form geoglyphs [8-13]. The geoglyphs seems to represent animals (may be, totemic animals), whose eyes are sometimes crated by artificial ponds. In a case, we see that the geoglyphs on the plain land are strongly connected with the terraced hills: in fact, we proposed the geoglyphs of Titicaca as an ancient graphic system based on artificial landforms [9]. Who is writing, A.C. Sparavigna, considers that the geoglyphs were created to mark the land of specific communities and that there was a strong connection between the agricultural system and the worship and burial places of Titicaca. These are personal conclusions coming from inspecting the satellite imagery of Google Maps [8]. It would be fundamental to have high resolution satellite images of all the basin, including the lagoons, to understand the extension of the agricultural system, and on-site researches to have the chronology of sites.

For what concerns the agricultural and meteorological studies, a quite interesting paper on the management of this system and on the physical process-based models is Ref.[14]: this paper proposed a model to explain the role played by the canals in the nocturnal heat dynamics and the cold mitigation process. This model consists of a two-layer transfer scheme with a vegetation layer and a substrate layer representing the canals. The calculations of Ref.14 show that the presence of a heat flux emanating from the canals and a corresponding water condensation on the crop, are both contributing to mitigate the environmental conditions, avoiding the crops to be frozen.

Another study [15] concerns the prehistoric diets, including analysis of stable isotope data from cooking pots, plants, animals and human teeth that have been collected by the Taraco Archaeological Project working in the Titicaca Basin of Bolivia. It is peculiar the analysis of the archaeological fish samples to understand their role in the diet of the Formative Period inhabitants of the southern Lake Titicaca Basin. According to the researchers, to understand the role of ichthyic resources in the human history, it is necessary to analyse the ecology of the fish from their muscle, bone and scales, since muscle is rarely preserved in archaeological contexts, whereas bone and scales are. For this reason, the researchers investigated the modern fish specimens from Lake Titicaca to compare with archaeological fish remains.

The physical modelling of this ancient agricultural system and the relevance of fishes in local diets, have to be considered in the framework of the paleoclimatic researches. This is important because the knowledge of the past climate (dry or wet) could help in evaluating the extension of the agricultural system and the amount of ichthyic resources of the lake.

In general, the study of lacustrine records is considered useful for understanding the mechanisms and effects of climate change. This is why Lake Titicaca is an important site for paleoclimatic research in the South American tropics because of the evidence for major lake level changes in the late Quaternary ([16], and references therein). The lake has an outlet, the Rio Desaguadero, but today, the most of the water is lost by the lake due to evaporation. This means that the lake is a nearly closed basin and this fact is increasing the sensitivity of the hydrologic mass balance of the lake to climate change. In [16], the research was performed by means of seismic stratigraphy: this analysis indicates that late-Quaternary lake levels have varied significantly, most likely because of climatic change. The seismic data used in conjunction with sediment core data indicate that there is a basin wide stratigraphic marker, most likely due to volcanic ashes.

According to Ref.17, South America has a scarcity of sites with century-scale paleoclimate data sets, but these data are extremely important because of the El Niño/Southern Oscillation events (ENSO), the migrations of the intertropical convergence zone (ITCZ) and the presence of the vast Amazon basin. According to [17], it is the Lake Titicaca drainage basin and its associated altiplano endorheic system, in particular the nearby alpine glaciers, that are containing important paleoclimate records. In [17], the researchers are describing a finely resolved record of lake-level change driven by climatic variability over the past 3500 years. The paper reports evidence that suggests a rapid lake-level rise of 15 to 20 m about 3500 years before present, and several century-scale low stands at 2900–2800, 2400–2200, 2000–1700, and 900–500 cal yr before present. These findings improve the knowledge of the timing, duration, and magnitude of variations in the precipitation–evaporation balance of the South American altiplano during the late Holocene. The study is based on radiocarbon chronologies necessary to resolve century-scale dynamics of precipitation–evaporation variations on the altiplano.

In Ref.18, researchers found two major dust events reaching maximum intensity at A.D. 600 and 920. They note that the dust could have been produced by the combination of extensive use of agricultural raised fields and the exposure of large areas of lake sediment during the periods of low-lake stands. According to [17], the peaks in dust content correspond with periods of major raised-field activity by the Tiwanaku civilization [19]. As reported in Ref.17, during an on-site travel in the years 1995 and 1996, researchers observed a several-meter decline in lake level that exposed very large areas of totora beds and lake sediment, that were quickly used for agricultural purposes. Time series of the yearly rise for the years 1915 to 1981 have been investigated: the relative spectral density clearly shows peaks with periods of 10.6 and 2.4 years [20]. Let us consider that the level of the lake is also oscillating during the year.

In Ref.21, it is claimed that the study of the past climatology of Peruvian altiplano demonstrated that the emergence of agriculture (ca. 1500 B.C.) and the collapse of the Tiwanaku civilization (ca. A.D. 1100) coincided with periods of abrupt and profound climate change. Archaeological evidence establishes spatial and temporal patterns of local agriculture. Prior to 1500 B.C., aridity in the altiplano precluded intensive agriculture. According to Ref.21, during a wet period from 1500 B.C. to A.D. 1100, the Tiwanaku civilization and its immediate predecessors created agricultural methods that stimulated the population growth, with corresponding large human settlements. A prolonged drier period (ca. A.D. 1100–1400) caused the decline of food production, the fields were abandoned and the cultural system collapsed.

An analogue detailed study of the Peruvian part of the Lake could be very important to understand the role of climate on the raise of Colla-Sillustani civilization and its connection with Inca civilization, and, of course of previous human settlements. Let us remember that human gatherers are found both North and South of Lima, Peru, as early as 8000 BC. Mountain civilizations were Kotosh (2000 BC), Tiwanaku-Huari 700AD, Collas-Sillustani (Titicaca Lake, 1100AD) and finally the Inca culture 1532 (AD): all these cultures had influences in the Inca culture, including the apparently autochthonous Titicaca Lake (Aymara-speaking) cultures [22]. The Aymara language is considered more ancient than the Inca language and has not only been found in the Titicaca Lake area but also in mountains close to Lima. Aymara-speaking people widespread throughout the Peruvian area: the Quechua language was imposed later by the Inca conquest and also by the Spanish conquerors. Aymara-speaking people were long ago established around Titicaca Lake area in the so called "Collao" area (see [22], and reference therein). According to [22], a tribe coming from this lake area set out for Cuzco, where they established, they spoke Quechua and were called "Inga" or "Inca" people.

In Fig.1, the behaviour of the level of the lake is shown as a function of time [23], we can see clearly the oscillation between dry and wet periods. Other studies on late Pleistocene/Holocene paleoclimates of the Bolivian Altipiano using the analysis of ostracod content, palynology, sedimentology and radiocarbon dating have been proposed [24].

Let us conclude with a discussion on satellite imagery again, connected with the dry and wet periodic behaviour of the local climate. As told in Ref.17, as the lake level declines, the soil is quickly used for agricultural purposes. In observing the Google Maps of the Umayo and Machacmarca Lagoons, Figures 2 and 3, after an image processing, we see that the surface, that is actually subsided under the water, was once covered by raised fields. The lagoons too were subjected to the dry-wet oscillation. As previously told, an analysis as in Ref.17 of the two lagoons could give information on prehistoric human settlements. Let us consider that Sillustani, the burial place of Collas, is a peninsula of the Umayo Lagoon [13]

Near the shore of the Titicaca Lake we see (Figure 4) an area densely covered by the earthworks of the raised fields. Again they look as geoglyphs. In Figure 5, details of two of these earthworks are shown. These images are coming from an area near the shore that can be seen due to the fact that the level of the lake is actually subsiding. Other satellite inspections, such as with radar or infrared detectors, could be of great help in detecting all the archaeological sites of this Peruvian region. A complete inspection with Google Maps is in any case necessary to have a total description of the raised fields and the related structure of canals and ponds. Besides the common destiny of any landform composed of fine-grained materials to become wide and flat relieves as a consequence of natural degradation processes, the earthworks of Titicaca are also subjected to the human action that can quickly destroy them.

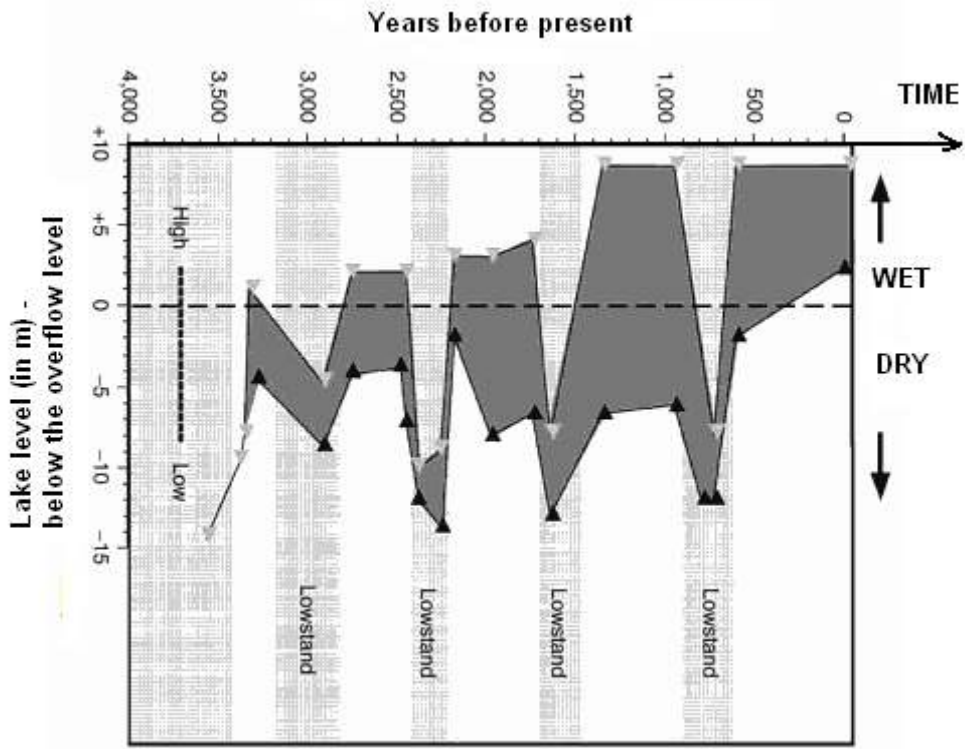

Figure 1: Level of the Titicaca Lake as a function of time. Image adapted from Ref.23.

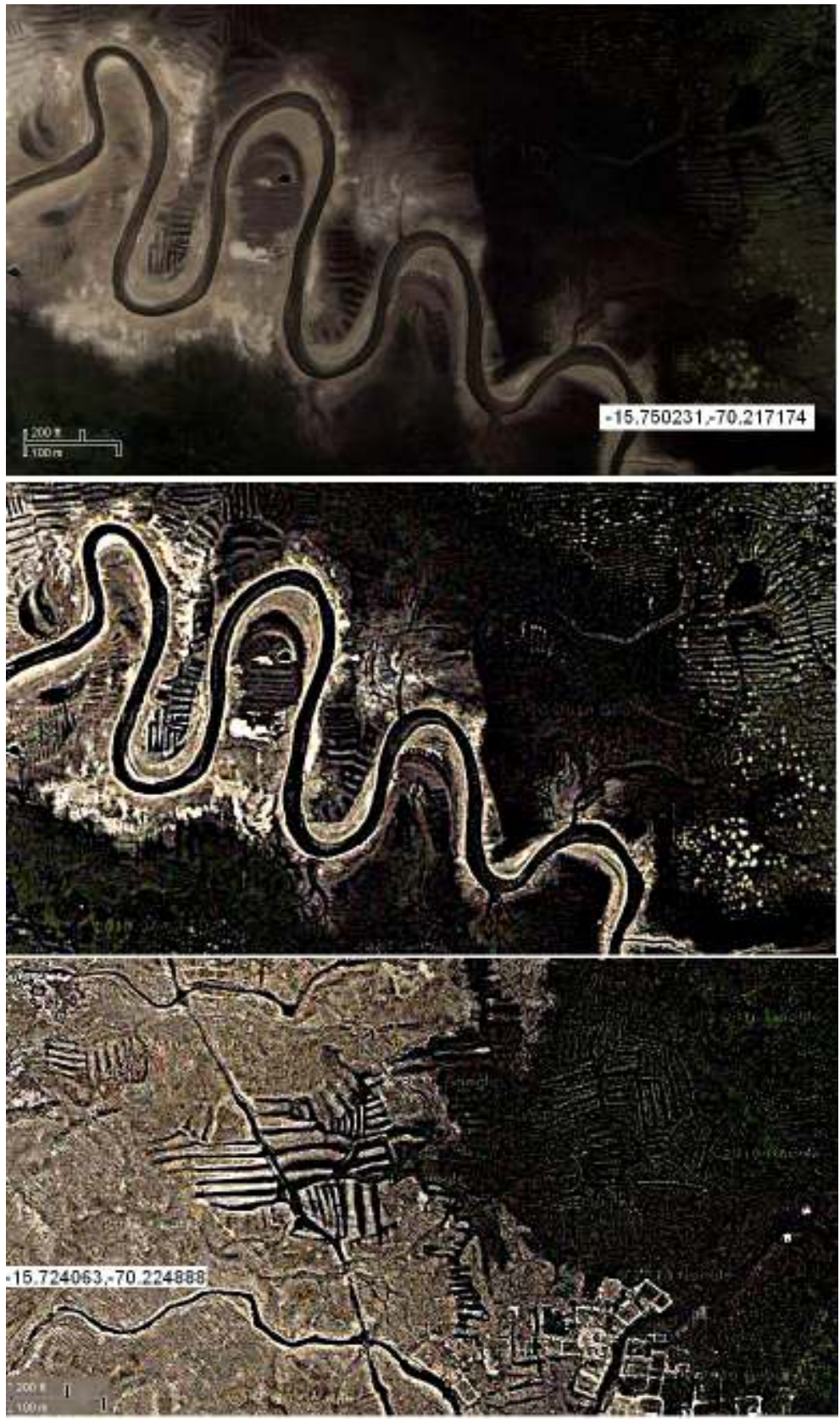

Figure 2: Two areas of the Umayo Lagoon as observed by Google Maps. After image processing (middle and low panels), we see that the level of the Lagoon was lower and that in a drier period, the exposed surface was cultivated with raised fields. Coordinates and scales are reported in the images.

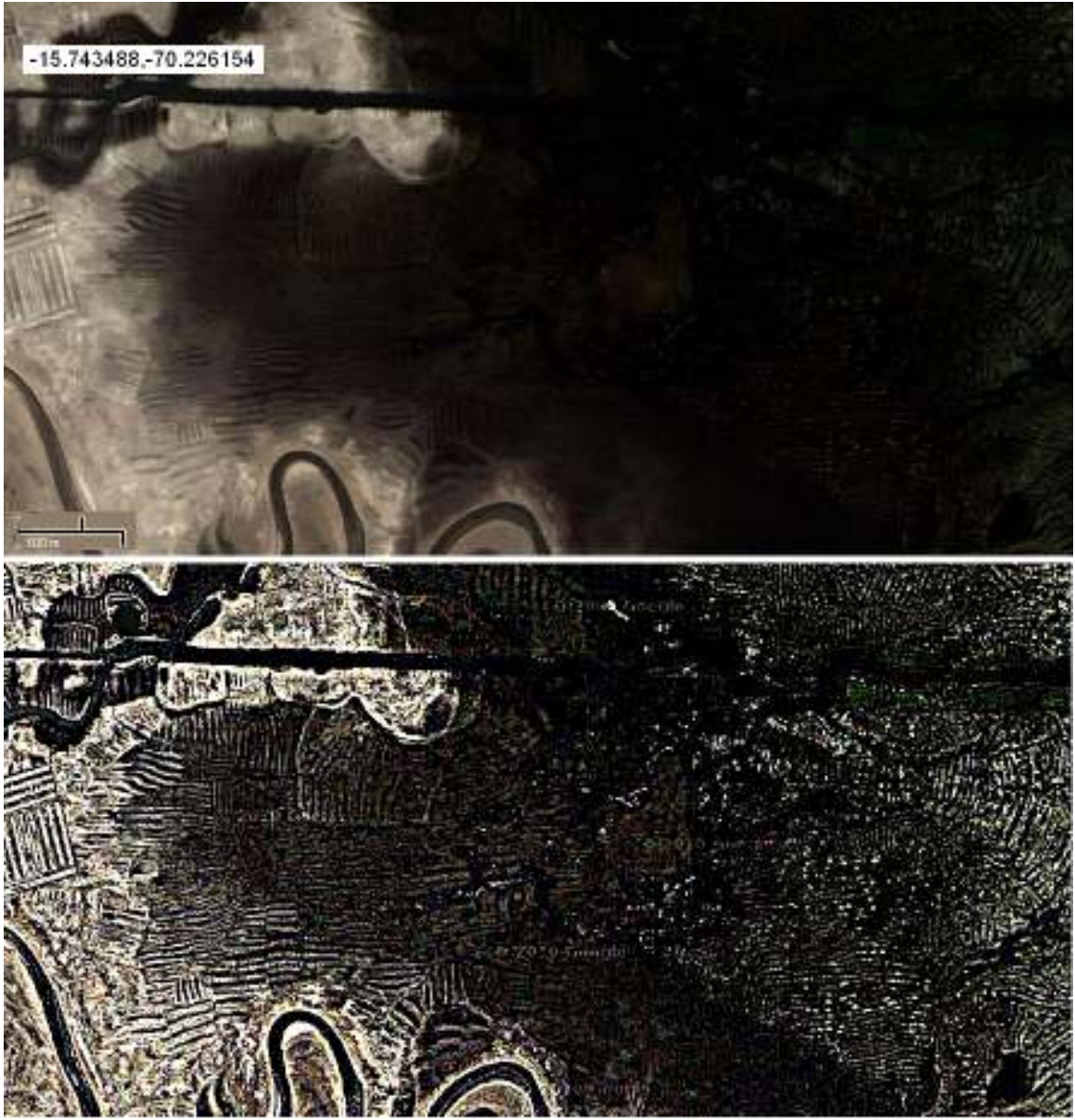

Figure 3: Another detail of Umayo Lagoon as observed by Google Maps. The lower panel shows the result after image processing. The level of the Lagoon was lower and the exposed surface was cultivated with raised fields. Coordinates and scale are reported in the image.

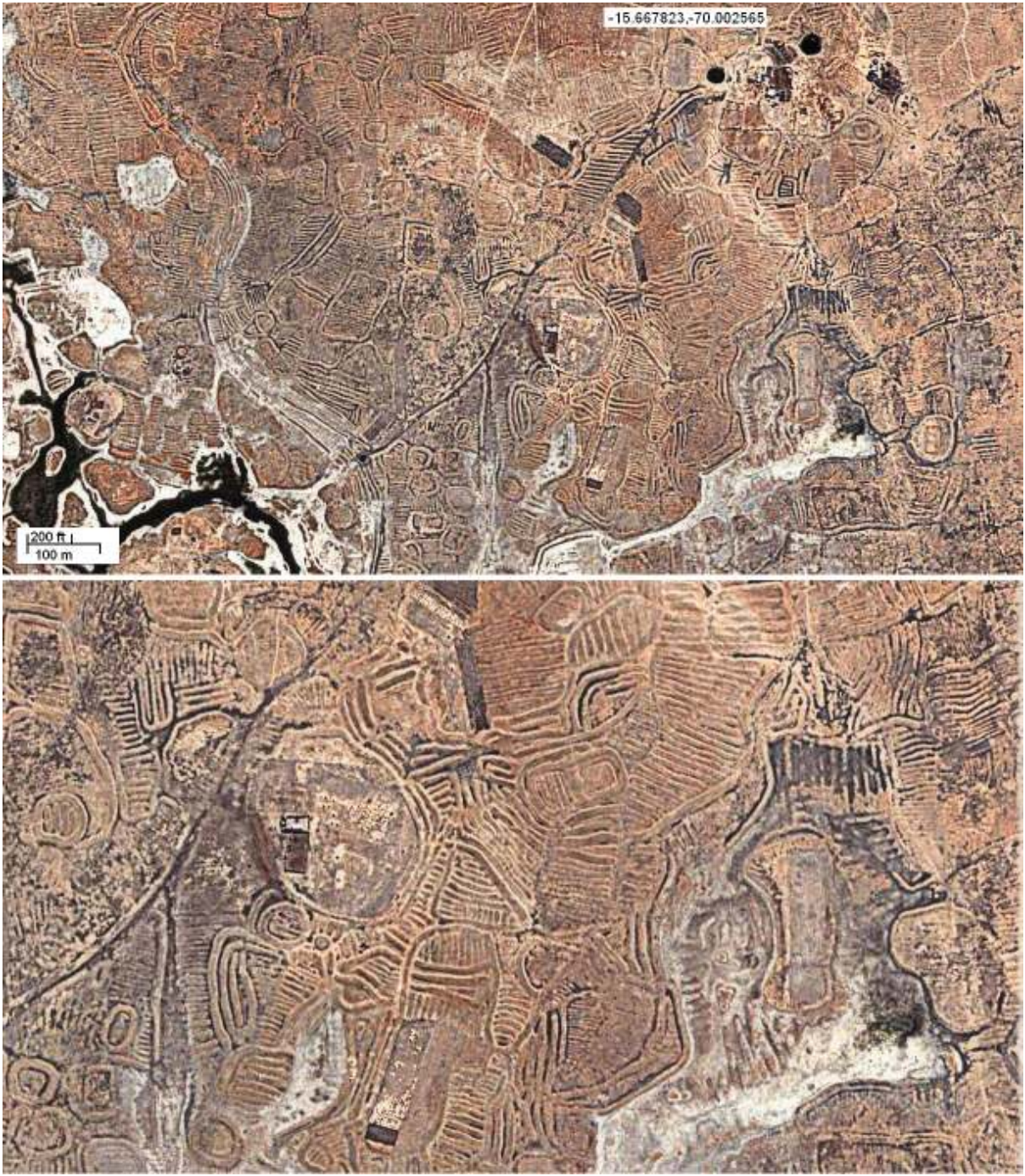

Figure 4: Near the shore of the Titicaca Lake we see an area densely covered by geoglyphs. Here a part of this region. This area can be seen due to the fact that the level of the lake is actually subsiding.

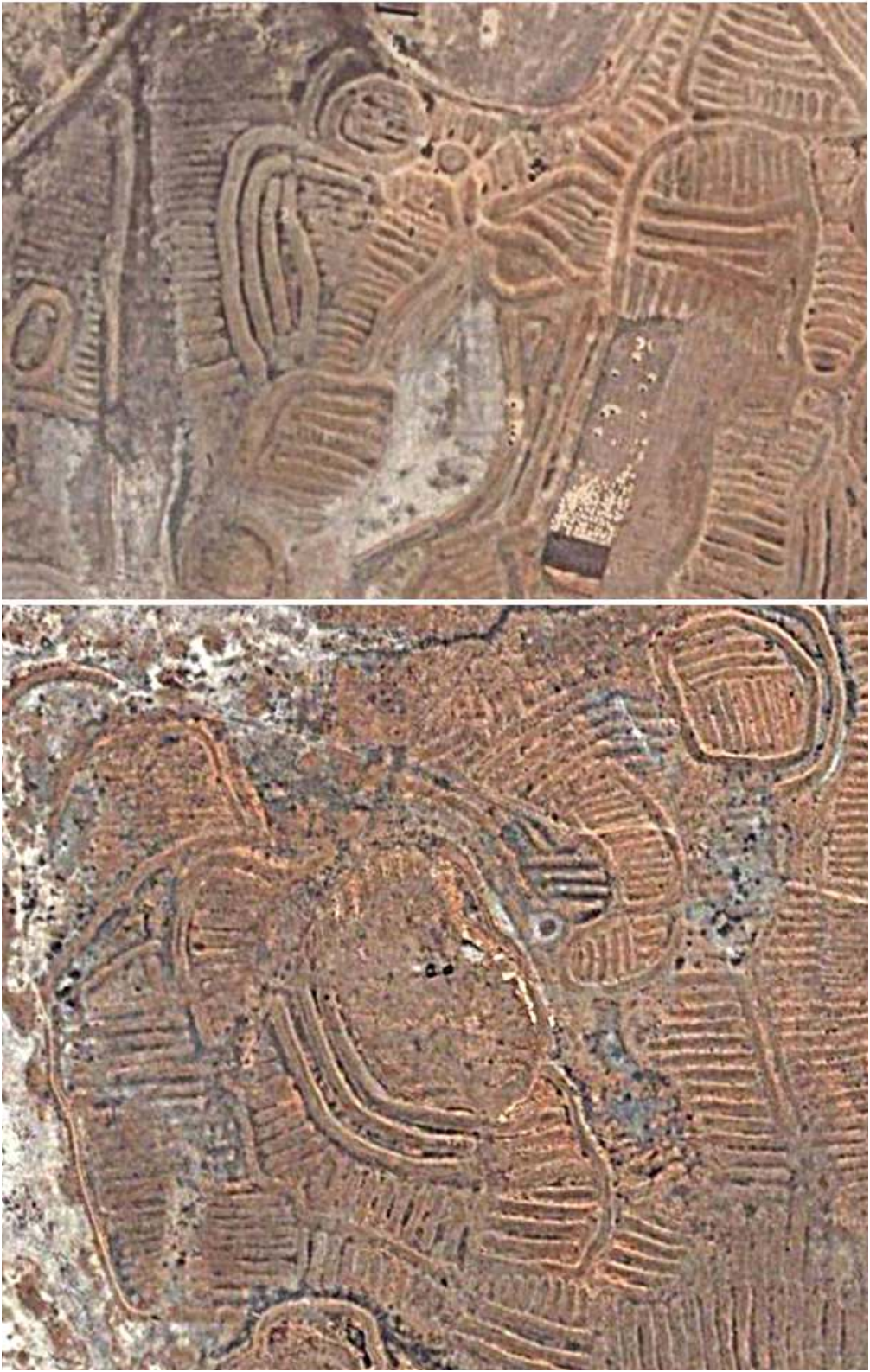

Figure 5: Two areas with earthworks creating figures on the ground. As told in Ref.17, as the lake level declines, the soil is quickly used for agricultural purposes. This is clearly observable in the upper panel, where it is possible to see the stacks of crops.